\documentclass[aps,prl,twocolumn,groupedaddress]{revtex4}
\begin{document}


\title{Bell's Theory with no Locality assumption.} 


\author{Charles Tresser}
\email[]{charlestresser@yahoo.com}
\affiliation{IBM, P.O.  Box 218, Yorktown Heights, NY 10598, U.S.A.}


\date{\today}

\begin{abstract}
We prove versions of the Bell and the GHZ Theorems that do not assume Locality but only the Effect After Cause Principle (EACP) according to which for any Lorentz observer the value of an observable cannot change because of an event that happens after the observable is measured. We show that the EACP is strictly weaker than Locality.  As a consequence  of our results, Locality cannot be considered as the common cause of the contradictions obtained in all versions of Bell's Theory.  All versions of Bell's Theorem assume Weak Realism  according to which the value of an observable is well defined whenever the measurement could be made and some measurement is made. As a consequence of our results, Weak Realism becomes the only hypothesis common to the contradictions obtained in all versions of Bell's Theory.  Usually, one avoids these contradictions by assuming Non-Locality; this would not help in our case since we do not assume Locality.  This work indicates that it is Weak Realism, not Locality, that needs to be negated to avoid contradictions in microscopic physics, at least if one refuses as false the de Broglie-Bohm Hidden Variable theory because of its essential violation of Lorentz invariance.
\end{abstract}


\maketitle

%
%
%
%

\noindent
\textbf{1)}  \textbf{Introduction.}
Following Bohm 's version \cite{Bohm} of the EPR \textit{gedanken experiment} \cite{EPR}, we consider entangled pairs of spin-$\frac{1}{2}$ particles such that the spin part of the wave function is the \emph{singlet state} (at any location pair $(x_1,x_2)$):
\begin{equation}\label{Singlet}
\Psi(x_1,x_2)=\frac{1}{\sqrt{2}}(| +\rangle _A\otimes| -\rangle_B-| -\rangle_A\otimes|
+\rangle_B)\,.
\end{equation}
This sum of tensor products represents an example of \emph{entanglement}, which means that this expression cannot be rewritten as one tensor product of one particle states.  We even have here a \emph{maximal entanglement} since all the summands have identical statistical weights.  The particles of the pair indexed by $i$ are called $(p_A)_i$ and $(p_B)_i$.  For each $i$ the particle $(p_A)_i$ flies to Alice who is equipped with the \emph{measurement tool} $E$ while $(p_B)_i$ flies to Bob who handles the \emph{measurement tool} $P$.  $E$ and $P$ can be chosen as Stern-Gerlach magnets if, following Stapp \cite{Stapp1971} we use  neutral particles, say neutrons.  There is a source  $S$ of entangled pairs and, together with the tools $E$ and $P$, the source $S$ is attached to the laboratory frame; we assume that the measurements at $E$ and $P$ are (essentially) simultaneous in that frame.  

Alice chooses the oriented axes $(a_A)_i$ and, with $ \textrm{Spin}(q)$ standing for the spin of particle $q$, she  observes the sequence $\mathcal E_i$ of normalized projections of $ \textrm{Spin}((p_A)_i)$ along $(a_A)_i$ while Bob chooses the oriented axes $(a_B)_i$ and observes the sequence $\mathcal P_i$ of normalized projections of $ \textrm{Spin}((p_B)_i)$ along $(a_B)_i$.  
Bohm \cite{Bohm} noticed in particular that any observation $\sigma \in \{-1,+1\}$ by Alice along $(a_A)_i$ would necessarily correspond, because of the structure of the singlet state, to the observation $-\sigma$ by Bob if he would choose $(a_B)_i=(a_A)_i$.  We will not recall nor revisit here the EPR paper, nor comment the way the authors themselves or Bell considered the issues raised in \cite{EPR} or in \cite{Bohm}.  The consideration of angles between the oriented axes  $(a_A)_i$ and  $(a_B)_i$ that can take any value in a setting that is otherwise the one proposed by Bohm is essential in the development of Bell's Theory \cite{Bell}, and it is precisely this theory that we revisit here (Bohm used right or zero angles between axes in \cite{Bohm} as he was merely proposing a new version of what he considered to be the content of the EPR paper \cite{EPR}).  The experiment that consists in emitting successive pairs in the singlet state and measuring the normalized projections of the associated spins is repeated a large number of times in order to get statistically significant results.  In order to achieve the same goal of significant statistics, the oriented axes $(a_A)_i$ and $(a_B)_i$ are usually kept constant for long sequences of values of $i$ (on such stretches of constancy, one may suppress the index $i$).  If the observation is $\mathcal Q_i$, we write $|\mathcal Q_i\rangle$ for the corresponding spin part of the state, and we denote by $X$ the sequence with generic element $X_i$. Thus, the correlation $\langle \mathcal U,\mathcal V\rangle$ is equal to Dirac's braket $\langle \mathcal U|\mathcal V\rangle$ whenever both  $|\mathcal U\rangle$ and  $|\mathcal V\rangle$ are quantum mechanical states, but we will prefer the statistical notation. We denote by $<a_1;a_2>$ the angle between the two oriented axes $a_1$ and $a_2 $ and we associate to any oriented axis $a_C$ the angle $\theta_C=<a_C;a_0>$  where $a_0$ is some oriented axis of  reference that points, say, horizontally and to the right when looking from the far side along the estimated classical trajectory of the departing particle flying toward Alice.  Recall that Quantum Mechanics predicts probabilities of equality or equivalently correlations, the equivalence of the two viewpoints being captured in our case in the following identity:
\begin{equation}\label{CorrelProba}
\langle  \mathcal E,  \mathcal P \rangle= 2 Prob( \mathcal E_i= \mathcal P_i)-1
\end{equation}
where $Prob($event$)$ is the probability of that $event$ (for a general reference for Quantum Mechanics covering somehow Bell's Theory and in particular the GHZ Theorem, see for instance \cite{Peres1993} or \cite{Le Bellac}).

\medskip
For the sequences $ \mathcal E$ and $ \mathcal P$ that we have defined for the  spin-$\frac{1}{2}$ singlet state (\ref{Singlet}), Quantum Mechanics predicts what we call the \emph{twisted Malus Law} that differs from the usual Malus law by the minus sign:
\begin{equation}\label{TwistedMalus}
\langle   \mathcal E,    \mathcal P \rangle= -\cos (\theta_A-\theta_B)\,.
\end{equation}
Since we only use spin-$\frac{1}{2}$ particles and normalized spin projections rather than photons and their polarization states, each time we mention in this paper the singlet state or Malus law (normal or twisted), we mean of course the spin-$\frac{1}{2}$ version of these objects (for a textbook presentation of both of the photons and the  spin-$\frac{1}{2}$ particles versions, see for instance \cite{Le Bellac}).  

\medskip
In the founding paper of Bell's Theory \cite{Bell} (see page 407 of that paper), Bell reached the conclusion that:

\smallskip
\emph{``In  a theory in which parameters are added to quantum mechanics to determine the results of individual measurements, without changing the statistical predictions, there must be a mechanism whereby the setting of one measuring device can influence the reading of another instrument, however remote.  Moreover, the signal involved must propagate instantaneously, so that such a theory could not be Lorentz invariant."}.

%

\bigskip
\noindent
More generally, the structure of a typical Bell type theorem reads either as the following statement that we call \emph{the Main Implication} or as its consequences as in Bell's citation just above:

\medskip
\noindent
\textit{ \,Quantum Mechanics} \quad \quad \quad \textit{Some inequality is violated }

\noindent
\textit{+ Augmentation choice} \,\,\,$ \Rightarrow \,\,$\textit{ for appropriate choices} 

\noindent
\textit{+ Extra hypothesis}\qquad \qquad \, \textit{ of some parameters.}

\medskip
\noindent
In the terms of the Main Implication, the example of ``Augmentation" chosen in Bell's 1964 paper \cite{Bell} is the assumption that there are ``Predictive Hidden Variables with the same statistics as Quantum Mechanics" while Bell's original example of ``Extra hypothesis" is ``Locality" that we next redefine both more formally and in such a way that the role of the augmentation be clearly stated.    

\medskip
\noindent
\textbf {Definition 1.}  \emph{Locality tells us that if $(x_0, t_0)$ and $(x_1, t_1)$ are spatially separated, \textit{i.e.,}  $\Delta x ^2> c^2 \Delta t^2$, then the setting of an instrument at  $(x_0, t_0)$ cannot change the output of a measurement made at  $(x_1, t_1)$. Furthermore, if one assumes that Weak Realism as defined below holds true,  the value of any observable that could be measured at $(x_1, t_1)$ in lieu of the observable that is actually being measured there is also independent of any instrument setting at $(x_0, t_0)$.}

\smallskip
\noindent
In the present paper we will show that the Extra hypothesis of Bell's Theorem can be chosen to be substantially weaker than Locality without affecting the truth of the Main Implication. 

\medskip
For a long time already many authors have proposed versions of Bell's Theorem based on Augmentations that are weaker than the Predictive Hidden Variables used in \cite{Bell}.  We recall that the concept of \emph{Predictive Hidden Variables} does not only mean that some variables make sense, even if beyond our reach, but that there are enough such variables so that using all variables, hidden or not, one would get a theory that would not only predict statistical results (like Quantum Mechanics) but would also predict the result of individual experiments and more generally of all the observables' values (even if  one cannot access these values). In particular all usual observables would have well defined values since they would be predictable, so that Predictive Hidden Variables, if they would exist, would imply \emph{Realism} (also called \emph{Microscopic Realism}, for instance in\cite{Leggett2008} ) in the sense that observables would have values independently of being observed or not.  We shall focus in this paper on an Augmentation of Quantum Mechanics that does not \emph{assume} any more predictive power than Quantum Mechanics; more precisely we shall only postulates that \emph{there is a value associated to any {\bf useful} measurement that could be made on a particle at the time when some measurement is made on that particle}.  This Augmentation is mostly Stapp's  \emph{Contrafactual Definiteness} \cite{Stapp1985} (see also \cite{Stapp1971}), also called \emph{Macroscopic Counterfactual Definiteness} and abbreviated as MCFD by Leggett who also offers an interesting discussion of it in \cite{Leggett2008}, is not only implied by the hypothesis used by Bell in \cite{Bell} but also by Microscopic Realism as explained in \cite{Leggett2008} and also by what is called sometimes \emph{the EPR  condition of reality} \cite{EPR}.  

\medskip
The Augmentation of Quantum Mechanics that we choose in this paper to develop a restricted hypotheses version of Bell's Theory is the small modification of MCFD that is stated above (where we have added the word {\bf ``useful"} in the usual definition of MCFD), is constructed as \emph{the weakest form of realism sufficient to develop Bell's Theory}  while (by a standard hypothesis going back to\cite{Bell} preserving the statistical predictions of Quantum Mechanics.  This is why so that we call it simply \emph{Weak Realism}. At first reading, the readers may as well choose the formulations of realism at the microscopic level that they like most as the definition of Realism, in lieu of our minimalist concept. 

\medskip
The concept that we introduce next is another essential ingredient of our work: it will be our Extra Hypothesis in the Main Implication, but we will mostly use it in a form that is much more specific that the one proposed here.  We have judged that it was preferable to stay at this less technical level in the Introduction, and we will also use the following high level description anyhow.    

\medskip
\noindent
\textbf {Effect After Cause Principle (\emph{EACP} - General Form):} \emph{- (i) For any Lorentz observer the value of an observable cannot change as a result of any cause that happens after said observable has been measured for that observer.}

\noindent
\emph{- (ii) Furthermore, if one assumes that Weak Realism holds true,  the value of any observable that could be measured at $(x, t)$ where some other observable is measured, but that is only inferred to exist at $(x,t)$ by invoking Weak Realism, cannot change as a result of any cause that happens after said non-observed observable gets a value at $(x,t)$ as a result of Weak Realism for that observer.}

\bigskip
The subtlety of the statement of the EACP calls for some special comments that are more of the ``warning" type than ordinary remarks.

\medskip 
\noindent
\textbf{Warnings.} \emph{
\begin{description}
\item[W1] We notice that time ordering makes sense in the definition of the EACP but that it is relative to the chosen Lorentz observer. 
\item[W2] It is crucial to point out that the value of an observable may well depend on a ``later" event (in the time ordering of the chosen Lorentz observer: see Warning [W1]) in the time structure of the chosen Lorentz observer: the next warning [W3] provides limits for the dependence on later events. 
\item[W3] A reading, once performed (or potentially performed when dealing in particular to entities that only exist if one assumes Weak Realism) cannot (further) change because of a later cause, but may depend on a later cause as when one assumes Non-Locality.
\end{description}
} 

After showing that the EACP is an hypothesis strictly weaker than Locality (the one line proof being that the EACP is compatible both with Locality and with Non-Locality) we will prove a version of Bell's Theorem where we only assume Weak Realism and  the EACP.  Let us recall that a typical formal statement of Bell's Theorem consists in the \textit{falsification of some inequality} (meaning as usual the exhibition of an instance such that the inequality that one attempts to \textit{falsify} indeed reduces to a false inequality between two numbers): such a falsification then lets one draw a conclusion as in Bell's citation reported above.  Not so surprisingly, there is a price to pay for the increased generality of the Bell's Theorem that we will prove in this paper. More specifically, in order to compensate for our weaker assumptions (that consist as usual in some choice of Augmentation and an Extra Hypothesis) the selection of an inequality and of its parameterization needs to be much more controlled than in former versions of Bell's Theory in order to produce a falsification of at least one of the Bell's Inequalities than what one needs when as usual on assumes Locality.  In particular, avoiding to assume Locality will not permit us to falsify  any inequality that uses four angles 
and more precisely 
two angles for each of the two particles of the singlet state as in the CHSH version of Bell's Theorem (see \cite{CHSH69, Bell71, CH74, AspectEtAl1982}). However the original configuration with three angles used in Bell's paper \cite{Bell} can be dealt with, but then only for some angle on the side where one uses two oriented axes, when one only assumes the EACP.  It might be the case that other configurations also work besides the one that we could find, but the fact that we cannot conclude using two angles for each of the two particles of the singlet state will turn out to be deeply linked to the difference between Locality and the EACP.

\medskip
We conclude from our results that \emph{the only common cause to all contradictions in Bell type theorems is whatever form of realism that one uses to augment Quantum Mechanics}.  In particular, invoking Non-Locality cannot prevent the contradictions that we establish.  Theorems of the type of Bell's can thus be used to strongly suggest that \emph{ Weak Realism is false}  (see also  \cite{ETP, Tresser2} and Remark 3 below).  Such a conclusion would be in line with the opinion that any form of realism at the microscopic level is in contradiction with the spirit of the Uncertainty Principle.  Rejecting as usual ``Local Realism", \textit{i.e.,} the conjunction of some form of realism and Locality, appears to be a misleading conclusion in view of our results. 

\bigskip
This paper is organized as follows. In Section 2 we complete the description, started above in this section, of the setting of the \emph{gedanken experiments} dealt with in Bell's Theory: for the sake of completeness, simple classical derivations of Bell's Inequality and Bell's Theorem, in the original  and CHSH form, are provided there assuming both Weak Realism and Locality as in the classical Bell's Theory.  In Section 3, we prove that the EACP is weaker an hypothesis than Locality.
We will also provide in Section 3 a computation of one special example of correlation that we call the \emph{No Correlation Lemma}.  The proof of our version of Bell's Theorem is then developed to conclude Section 3. 

\bigskip
\noindent
\textbf{2)}  \textbf{Setting, statement and Proofs in usual Bell's Theory.}
We started the description of the experiment in the Introduction, but what was described there of the EPR-Bohm \textit{gedanken experiment} is not yet enough to reach any form of Bell's Theorem. The Uncertainty Principle \cite{Heisenberg} tells us that only one spin projection, \textit{i.e.,} one axis, can be chosen for each of the two particles $p_A$ and $p_B$, not enough to generate any meaningful inequality relating different correlations. In order to get enough observables to build a meaningful inequality, one needs to ``augment" Quantum Mechanics into a candidate for a theory of microphysics that would coincide with Quantum Mechanics where Quantum Mechanics has something to tell us, and that is compatible with the statistical predictions of Quantum Mechanics (which have been proven right by numerous experiments over the years).  Using the Augmentation of Quantum Mechanics by any form of realism to have more values of observables at once necessarily turns what we started to describe as an experiment into a \textit{gedanken} experiment.  Of course, the legitimacy of such an augmentation of Quantum Mechanics is questionable and we hope that we help to make the case (see Remark 3) that indeed,  Weak Realism violates the laws of Physics.  But this will not prevent us from assuming often  Weak Realism as we argue \textit{ad absurdum}.  

\medskip
The following two conventions are adopted in a more or less explicit form in all works on Bell's Theory, independently of the strength of the Augmentation being chosen:

\bigskip
\noindent
\textbf{Convention 1.}  \emph{Whenever we assume that Quantum Mechanics is augmented by a form of realism, we implicitly postulate that \emph{any quantity that is not measured but that exists according to the Augmentation has the value that would have been measured if this quantity would have been the one measured, the world being otherwise unchanged}.  It seems to us that the meaning of the value of an observable makes no much sense otherwise so that this convention is probably the most consensual component of this paper. }

\medskip
\noindent
\textbf{Convention 2.}  \emph{Whenever we assume that Quantum Mechanics is augmented by a form of realism, we assume that said Augmentation is made \emph{without changing the statistical predictions}.  This is (up to wording) the assumption that Bell made in his foundational 1964 paper \cite{Bell}, except for the fact that we do not restrict the choice of Augmentation to Predictive Hidden Variables.}

\medskip
As we shall see, Convention 2 is not enough to get convergence, nor even evaluation for averages over finite sums for all the correlations that we need.  Historically, versions of inequalities involving either three sequences of spin projections (what we call \textit{``version $V3$"}) or four sequences of spin projections (what we call \textit{``version $V4$"}) have been used, and it will be important for us to use both versions.   More precisely, we will use: 

- Version $V3$ in order to get our Bell Theorem without Locality  in subsection 3.4.  

- Versions $V3$ and $V4$ in order to illustrate why the EACP is a weaker hypothesis  than Locality in subsection 3.2.
\noindent
We will also use both versions $V3$ and $V4$ to examine closely in subsection 3.2 what would be the cost of abandoning Locality when dealing with the usual Bell's Theory that uses Locality as an essential assumption.  

\medskip
Coming back to the setting and notations of the Introduction, and assuming Weak Realism so that extra axes $(a_A)'_i$ and $(a_B)'_i$ can respectively be chosen by Alice and Bob, we end up having at our disposal the following sequences:  

- The sequences $\mathcal E_i$ on Alice's side using axes $(a_A)_i$ and  $\mathcal P_i$ on Bob's side using axes $(a_B)_i$ are the two sequences of normalized spin projections that are actually observed. 

-  The sequences $\mathcal E'_i$ on Alice's side and  $\mathcal P'_i$ on Bob's side that are the two sequences of what would be supplementary normalized spin projections, with values that are most probably out of reach. Such supplementary normalized spin projections values would be well defined - even if out of possible knowledge - if and only if  Weak Realism or some stronger form of Microscopic Realism holds true (justifying in part the statement that Weak Realism is the weakest form of Realism that can be used to develop Bell's theory).  These sequences are supposedly what one would get respectively along the axes  $(a_A)'_i$ and $(a_B)'_i$ if those axes would be used to measure normalized spin projections instead of the axes $(a_A)_i$ and $(a_B)_i$.  Even if such supplementary sequences of normalized spin projections cannot be known, one may construct out of them some objects with statistical significance such as correlations or probabilities of equality on which one has grip under the standing assumption that whichever form of Microscopic Realism that one invokes must respect the statistical predictions of Quantum Mechanics.  

\medskip
One may think that the range of the index $i$ is cut into disjoint intervals $I_\kappa$ so that for any $I_\kappa$  the axes $(a_A)_i, (a_B)_i, (a_A)'_i, (a_B)'_i$ do not vary  with $i$ as long as $i$ stays in $I_\kappa$: we shall denote by $N_\kappa$ the number of elements of  $I_\kappa$. All the sequences that we have introduced are sequences of normalized spin projections for spin-$\frac{1}{2}$ particles, hence sequences of $-1$'s and $1$'s. We shall next focus on abstract sequences and finite chunks of sequences of $1$'s and $-1$'s.  

\bigskip
\noindent
\textbf{2.1: The formal aspects of Bell's Inequalities:}
We now follow Sica \cite{Sica1, Sica2} (except that we defer deciding which quantities need a prime) who noticed that if $w_i$, $x_i$, $y_i$ and $z_i$ are four sequences with values in the set $\{-1,1\}$, then one has simple factorization identities that  lead \textit{via} simple algebra to inequalities involving either three or four sequences or finite chunks of these sequences. For version $V3$, we use the fact that $y_i^2\equiv 1$ to start with:

\begin{equation}\label{Sica_3_1}
x_iy_i-x_iz_i=x_iy_i(1-y_iz_i)
\end{equation}
so that by summing over the elements of  $I_\kappa$, dividing by  $N_\kappa$ and taking absolute values, we get:
\begin{widetext}
\begin{equation}\label{Sica_3_2}
|\sum_{i\in I_\kappa}\frac{x_iy_i}{ N_\kappa}-\sum_{i\in I_\kappa}\frac{x_iz_i}{ N_\kappa}|\leq\sum_{i\in I_\kappa}\frac{|x_iy_i|\cdot|1-y_iz_i|}{ N_\kappa}\leq 1-\sum_{i\in I_\kappa}\frac{y_iz_i}{N_\kappa}\,.
\end{equation}
 \end{widetext}
 Thus 
\begin{equation}\label{Sica_3_3}
|\sum_{i\in I_\kappa}\frac{x_iy_i}{ N_\kappa}-\sum_{i\in I_\kappa}\frac{x_iz_i}{ N_\kappa}|\leq 1-\sum_{i\in I_\kappa}\frac{y_iz_i}{N_\kappa}\,.
\end{equation}
Assume then that there is convergence as $ N_\kappa\to\infty$.  Denoting by $\langle f,g\rangle$ the correlation of two functions $f$ and $g$, we get: 
\begin{equation}\label{Sica_3_4}
| \langle x,y\rangle -\langle x,z\rangle|\leq 1-\langle y,z\rangle\,,
\end{equation}
one formal form of the $V3$ version of Bell's Inequalities. 

\smallskip
We now turn to the algebra of the version $V4$. Again following Sica we start with:

\begin{equation}\label{Sica_4_1_a}
x_iy_i+x_iz_i+ w_iy_i-w_iz_i=x_i(y_i+z_i)+w_i(y_i-z_i)\,.
\end{equation}

\bigskip
\noindent
Simple manipulations on this identity then yield:
\begin{widetext}
\begin{equation}\label{Sica_4_2}
|\frac{1}{N_\kappa}\sum_{I_\kappa}x_iy_i+\frac{1}{N_\kappa}\sum_{I_\kappa}x_iz_i|
+
|\frac{1}{N_\kappa}\sum_{I_\kappa}w_iy_i-\frac{1}{N_\kappa}\sum_{I_\kappa}w_iz_i|
\leq
\frac{1}{N_\kappa}\sum_{I_\kappa}|x_i|\cdot|y_i+z_i|+\frac{1}{N_\kappa}\sum_{I_\kappa}|w_i|\cdot|y_i-z_i|
\end{equation}

Now, since $\min(|y_i+z_i|, |y_i-z_i|)=0$ and  $\max(|y_i+z_i|, |y_i-z_i|)=2$, equation (\ref{Sica_4_2}) can be rewritten as

\begin{equation}\label{Sica_4_3}
|\frac{1}{N_\kappa}\sum_{I_\kappa}x_iy_i+\frac{1}{N_\kappa}\sum_{I_\kappa}x_iz_i|
+
|\frac{1}{N_\kappa}\sum_{I_\kappa}w_iy_i-\frac{1}{N_\kappa}\sum_{I_\kappa}w_iz_i|
\leq 2
\end{equation}
\end{widetext}
Assuming convergence, the averages generate correlations and one obtains the following form of the CHSH inequality, our $V4$ version of Bell's Inequalities
\begin{equation}\label{Sica_4_4}
 |\langle x,y\rangle +\langle x,z\rangle| + |\langle w,y\rangle -\langle w,z\rangle|  \leq 2\,,
\end{equation}
which contains equation (\ref{Sica_3_4}) as a special case (first restrict to $x=y$, replace each $x$ by $y$ and then rename $w$ to $x$). 
We notice that when two sequences are actually observed so that elements with the same index come from the same pair,
then Quantum Mechanics provides the value of the correlation and in particular guaranties convergence. Be it in version $V3$ or version $V4$, we made no attempt to deduce all the Bell's inequalities, formal or not. For that and the statistical aspects of Bell's Inequalities and Bell's Inequalities as a particular case of Boole's Inequalities, see for instance \cite{Fine1982a, Fine1982b, Boole1862, Pitowsky1989a, Pitowsky1991, Pitowsky1994, Pitowsky2001}.

\bigskip
\noindent
\textbf{2.2: From formal inequalities to Bell's Inequalities and Bell's Theorem:}
We have obtained versions $V3$ and $V4$ of Bell's Inequalities using abstract sequences of $1$'s and $-1$'s.  In order to come one step closer to physics, we first appropriately pair: 

- the symbols $\mathcal E_i$,  $\mathcal P_i$ that represent actual observations, 

- and the symbols $\mathcal E'_i$ and  $\mathcal P'_i$ that represent values provided by the Weak Realism assumption 

\noindent
to the sequences  $w_i$, $x_i$, $y_i$, $z_i$ used in deriving the inequalities (\ref{Sica_3_3}) and  (\ref{Sica_4_3}).  

\medskip
For version $V3$, we need to take $x_i$ and $z_i$ on the same side, \textit{e.g.,} Alice's side: thus $x_i=\mathcal E_i $ and  $z_i=\mathcal E'_i $, whence $y_i=\mathcal P_i $. Then equations (\ref{Sica_3_3}) and  (\ref{Sica_3_4}) become respectively:

\begin{equation}\label{Sica_3p_3}
|\sum_{i\in I_\kappa}\frac{\mathcal E_i\mathcal P_i}{ N_\kappa}-\sum_{i\in I_\kappa}\frac{\mathcal E_i\mathcal E'_i}{ N_\kappa}|\leq 1-\sum_{i\in I_\kappa}\frac{\mathcal P_i\mathcal E'_i}{N_\kappa}\,.
\end{equation}
and
\begin{equation}\label{Sica_3p_4}
| \langle \mathcal E,\mathcal P\rangle -\langle \mathcal E,\mathcal E'\rangle|\leq 1-\langle \mathcal P,\mathcal E'\rangle\,,
\end{equation}

\medskip
As for version $V4$, we want $w$ and $z$ to be the values generated by the  Weak Realism hypothesis, but  we need also $x_i$ and $z_i$ to be on differente sides and  $y_i$ and $w_i$ to be on differente sides. One way to achieve that is to choose the replacements
$x\to  \mathcal E$, $y\to  \mathcal P$, $w\to  \mathcal E'$, $z\to  \mathcal P'$. Thus equations (\ref{Sica_4_3}) and  (\ref{Sica_4_4}) become respectively:
\begin{widetext}
\begin{equation}\label{Sica_4p_3}
|\frac{1}{N_\kappa}\sum_{I_\kappa}\mathcal E_i\mathcal P_i+\frac{1}{N_\kappa}\sum_{I_\kappa}\mathcal E_i\mathcal P'_i|
+
|\frac{1}{N_\kappa}\sum_{I_\kappa}\mathcal E'_i\mathcal P_i-\frac{1}{N_\kappa}\sum_{I_\kappa}\mathcal E'_i\mathcal P'_i|
\leq 2
\end{equation}
\end{widetext}
and the following form of the CHSH inequality:
\begin{equation}\label{Sica_4p_4}
 |\langle \mathcal E,\mathcal P\rangle +\langle \mathcal E,\mathcal P'\rangle| + |\langle \mathcal E',\mathcal P\rangle -\langle \mathcal E',\mathcal P'\rangle|  \leq 2\,.
\end{equation}
Our main goal in this section is only to reach the classical Bell's Inequalities and Bell's Theorems under the usual hypothesis. We also want to inspect here the correlations that can be computed if one assumes Weak Realism and Locality. This examination of what is computable under these hypotheses will be used in the next section to compare the strengths of different hypotheses.

\medskip
We have already invoked  Weak Realism in order to give meaning to three spin projections at once in version $V3$, or four spin projections at once in version $V4$. In order to give meaning to the correlations in equations (\ref{Sica_3p_4}) and (\ref{Sica_4p_4}), we now further assume Locality, so that the sequences on one side do not depend on the choice of the axes along which the spin is projected on the other side. Then, under Conventions 1 and 2 that are both triggered by assuming Weak Realism, we can use the twisted Malus law, that gives us:
\begin{equation}\label{TMalus 2}
\langle \mathcal E, \mathcal P \rangle= -\cos (\theta_\mathcal E-\theta_\mathcal P)\,,
\end{equation}
to also obtain readily:
\begin{equation}\label{TMalus 3_1}
\langle \mathcal E, \mathcal P' \rangle= -\cos (\theta_\mathcal E-\theta_{\mathcal P'})
\end{equation}
and 
\begin{equation}\label{TMalus 3_2}
\langle \mathcal E', \mathcal P \rangle= -\cos (\theta_{\mathcal E'}-\theta_\mathcal P)\,.
\end{equation}
Let $\tilde{\mathcal Q}$ stand for the sequence or normalized spin projections along the angle $\theta_\mathcal Q$ but on the side opposite to the side corresponding to $\mathcal Q$.  Since in this subsection we are assuming Locality, we have the identity:
\begin{equation}\label{Conserv}
\tilde{\mathcal Q}_i+{\mathcal Q}_i\equiv 0
\end{equation}
for any  ${\mathcal Q}\in \{ \mathcal E, \mathcal P,  \mathcal E', \mathcal P'  \} $. The relation (\ref{Conserv}) is a direct consequence of the singlet state expression and Wave Packet Reduction if one at least of ${\mathcal Q}$ and $\tilde{\mathcal Q}$ is actually measured. For the other cases, one uses Locality to state that $\tilde{\mathcal Q}_i$ is unchanged if the setting is changed on the other side, where one could actually measure ${\mathcal Q}$.  But then, we notice that by Convention 1 and Locality, ${\mathcal Q}_i$ remains unchanged if it is measured instead of being inferred to make sense by invoking Weak Realism, so that in all cases the conclusion is the same as if one at least of ${\mathcal Q}$ and $\tilde{\mathcal Q}$ is observed. 

\medskip
We notice that if one does not assume Locality, then the identity (\ref{Conserv}) holds true when at least one of  ${\mathcal Q}$ and $\tilde{\mathcal Q}$ is actually observed, but not necessarily otherwise since one cannot then use the reasoning on which we relied to justify the relation (\ref{Conserv}) in the case when Locality is assumed to hold true. This difference between what can be deduced depending on whether one assumes or not Locality to hold true will be very important in the next Section.  

From equations(\ref{TMalus 3_1}) or (\ref{TMalus 3_2}) that are equivalent to each other by exchanging the sides of Alice and Bob,  we get readily:
\begin{equation}\label{TMalus 3_3}
\langle \mathcal E, \tilde{\mathcal E'}\rangle= -\cos (\theta_\mathcal E-\theta_{\mathcal E'})
\end{equation}
and 
\begin{equation}\label{TMalus 3_4}
\langle \tilde{\mathcal P'}, \mathcal P \rangle= -\cos (\theta_{\mathcal P'}-\theta_\mathcal P)\,,
\end{equation}
from which by (\ref{Conserv}) we respectively get:
\begin{equation}\label{TMalus 3_3_a}
\langle \mathcal E, {\mathcal E'}\rangle= \cos (\theta_\mathcal E-\theta_{\mathcal E'})
\end{equation}
and 
\begin{equation}\label{TMalus 3_4_a}
\langle {\mathcal P'}, \mathcal P \rangle= \cos (\theta_{\mathcal P'}-\theta_\mathcal P)\,.
\end{equation}
Using again (\ref{TMalus 3_3_a}), (\ref{TMalus 3_4_a}), and Locality, we also know that:
\begin{equation}\label{TMalus 3_5_a}
\langle \mathcal E',\tilde {\mathcal P'}\rangle= \cos (\theta_{\mathcal E'}-\theta_{\mathcal P'})
\end{equation}
and 
\begin{equation}\label{TMalus 3_5_b}
\langle {\mathcal P'},\tilde{\mathcal E'} \rangle= \cos (\theta_{\mathcal P'}-\theta_{\mathcal E'} )\,.
\end{equation}
Any of these two equations lets us compute $\langle {\mathcal P'},{\mathcal E'} \rangle$ as:
\begin{equation}\label{TMalus 3_6}
\langle \mathcal E',{\mathcal P'}\rangle= -\cos (\theta_{\mathcal E'}-\theta_{\mathcal P'})\,.
\end{equation}

We now have the values, hence also in particular the convergence of the finite sums as the numbers $N_\kappa$ diverge, for all the correlations that we need in both versions $V3$ and $V4$. Thus both of the Bell's Inequalities, \textit{i.e.,} equations (\ref{Sica_3p_4}) and  (\ref{Sica_4p_4}), that we have formally deduced assuming convergence are fully justified if one assumes Weak Realism and Locality. In order to get from Bell's Inequalities to Bell's Theorem one needs to falsify at least one of these inequalities by choosing appropriate values of the parameters (the oriented axes or equivalently the angles). We will provide falsifications for both versions $V3$ and $V4$.

\medskip
- For version $V3$ we choose $\theta_{\mathcal P}=0$, $\theta_{\mathcal E}=\frac{3\pi}{4}$, and   $\theta_{\mathcal E'}=\frac{-3\pi}{4}$ so that  $\theta_{\mathcal E}$ and  $\theta_{\mathcal E'}$ differ by a right angle.  Since using Locality we easily get $\langle \mathcal E,\mathcal E'\rangle=0$, by further using $\langle \mathcal E,\mathcal P\rangle =\langle \mathcal E',\mathcal P\rangle=\frac{\sqrt{2}}{2}$, and by replacing all the correlations in equation (\ref{Sica_3p_4}) by their respective values we end up deducing the false inequality $\sqrt{2}<1$ by specialization of the $V3$ version of Bell's Inequalities. We can thus conclude that at least one of the assumptions that we have made,  Weak Realism and Locality, must be a violation of the laws of microphysics. Many other choices of angles would also work to generate a falsification of equation (\ref{Sica_3p_4}) or another Bell's Inequality. \textbf{Q.E.D.}

\medskip
- For version $V4$ we choose $\theta_{\mathcal E}=\frac{\pi}{4}$, $\theta_{\mathcal E'}=\frac{3\pi}{4}$, $\theta_{\mathcal P}=\frac{\pi}{2}$, and $\theta_{\mathcal P'}=0$, thus angular differences $| \theta_{ \mathcal E}- \theta_{ \mathcal P}|=| \theta_{ \mathcal E}-\theta_{ \mathcal P'}|=| \theta_{ \mathcal E'}-\theta_{ \mathcal P}|=\frac{\pi}{4}$ and $|\theta_{  \mathcal E'}-\theta_{ \mathcal P'}| =\frac{3\pi}{4}$.  Replacing the correlations in equation  (\ref{Sica_4p_4}) by their respective values we end up having deduced the false inequality $2\sqrt{2}\leq 2$ by specialization of the $V4$ version of Bell's Inequalities. Thus we can  again conclude that at least one of the assumptions that we have made,  Weak Realism and Locality, must be a violations of the laws of microphysics. Many other choices of angles would also work here, but the example chosen here for $V4$ is optimal in terms of the worse falsification of (\ref{Sica_4p_4}).  \textbf{Q.E.D.}

\medskip
\noindent
\textbf{Remark 1.} \emph{As Sica noticed in \cite{Sica1}, the finite $N_\kappa$ equations  (\ref{Sica_3p_3}) and (\ref{Sica_4p_3}) are identities, independently of any convergence property.  Sica calls them \emph{``Bell Identities"} to distinguish them form the \emph{``Bell Inequalities"} that follow from these identities if convergence hold true for all the averages.  The Bell Identities have to be satisfied as soon as one assumes Weak Realism that provides us the three or four sequences of $-1$'s and $1$ that are needed depending on which of these two identities we want to work with. Nevertheless, it would at best hard to use the Bell Identities to get a contradiction if one had no proof of convergence since then one would not have means to evaluate the terms in the identities (whether one deals with finite sums or with their asymptotic values).} 

\bigskip
\noindent
\textbf{3)}  \textbf{A Bell's Theorem with no Locality assumption.}
In subsection 2.2, we have recalled the classical theory of Bell, in two versions $V3$ and $V4$ where the number in the name of the version is the number of oriented axes used to obtain normalized projections of the spins. We completed this task assuming  Weak Realism and Locality. 

\medskip
\emph{But what happens if Locality is replaced by the EACP?} 

\medskip
\noindent
We will first investigate what remains of the computability of the various correlations that are related by one or another Bell Inequality. We will see that only $V3$ can be dealt with when assuming the EACP instead of Locality: one of the correlations in equation (\ref{Sica_4p_4}) cannot be evaluated, nor even guaranteed convergence with the substitute hypothesis. 

\bigskip
\noindent
\textbf{3.1:}  \textbf{Statement of the EACP vs Locality Lemma.}
We will use, here and in the next subsection, a version of the Effect After Cause Principle that is quite focused on the entities that we deal with in Bell's Theories. 

\bigskip
\noindent
\textbf {Effect After Cause Principle (\emph{EACP}).} \emph{For any Lorentz observer and for any $\mathcal Q$ in $\{ \mathcal E,\, \mathcal E',\, \mathcal P ,\, \mathcal P'  \}$, a value $\mathcal Q_i$ of $\mathcal Q$ cannot change as a result of a cause that happens after $\mathcal Q_i$ has been measured for that observer.}

\medskip
This version of the EACP adapted to the context of Bell's Theory will be used to prove the following result that is crucial to our purpose, but the reader is advised that warnings [W1]-[W3] are still in vigor and are essential for a correct understanding of any of the definition of the EACP, be it at high level as before or down to earth as in the version just stated.

\medskip
\noindent
\textbf{EACP vs Locality Lemma.} \emph{The EACP is different from Locality, and in fact strictly weaker than Locality.}  

\smallskip
\noindent
Otherwise speaking, Locality implies the EACP but the reverse implication is not true.

\medskip
Before proving this lemma, we give a definition that will be useful whenever dealing with the EACP throughout the rest of the paper.

\medskip
\noindent
\textbf{Definition 2.}  \emph{With $(X,Y)\in \{(E,P), (P,E)\}$ an \emph{$X$-$Y$ observer} is a Lorentz observer for whom measurements at the measurement tool $X$ occur before measurements at measurement tool $Y$ for each pair produced at $S$.}

\bigskip
\noindent
\textbf{3.2:}  \textbf{Proofs of the EACP vs Locality Lemma.}

\smallskip
\noindent
\emph{First proof of  the EACP vs Locality Lemma.} We assume  Weak Realism and the EACP and notice that, in view of Warning [W1]-[W3], the EACP has been formulated so as to be compatible either with Locality or with Non-Locality.  \textbf{Q.E.D.}

In particular, the correlation  $\langle \mathcal E',\mathcal P'\rangle$ does not make sense as long as one does not \textbf{also} assume Locality or anyway some strong enough extra condition on top of the EACP, so that version $V4$ of the Bell inequality cannot be used if one replaces assuming Locality by assuming the EACP only. The status of $V3$ is different enough and we will come back to the problem of the computation of  $\langle \mathcal E,\mathcal E'\rangle$  or $\langle \mathcal P,\mathcal P'\rangle$ under the EACP assumption in some particular cases in the proof of what we call The No-Correlation Lemma below. To the contrary, we notice that formulas \ref{TMalus 2} to \ref{TMalus 3_2} remains valid when assuming only the EACP, together with the convergence properties embedded in these formulas.

\bigskip
\noindent
\emph{Second proof of  the EACP vs Locality Lemma.}
We aim here at the same lemma but for the EACP as it appears in its general definition given in the Introduction. For that, we first notice that the EACP is nothing but Causality in a world without augmentation of Quantum Mechanics by any form of realism.  On the other hand Causality is known to be the impossibility of signaling. Since one knows that the violation of Locality, independently of realism,  does not permit signaling (see, \textit{e.g.,} \cite{Eberhard1978}, cite{GRW1980}, \cite{Jordan1983}), we know that Locality and Causality do not coincide in a world without augmentation of Quantum Mechanics by any form of realism: indeed Locality is stronger that Causality in such a world. We deduce that the EACP is not Locality, and is in fact weaker than Locality in a world without augmentation of Quantum Mechanics by any form of realism. 

While we aim at proving that we are indeed in a world without augmentation of Quantum Mechanics by any form of realism, we cannot use this fact and need to make sure of what happens if one assumes that Weak Realism holds true.  Then the negation of the EACP for non-observed observables is in conflict with Convention 1 so that, like in the case when one does not assume that Weak Realism holds true, the negation of the EACP can only have a chance to hold true if Causality Fails. Thus, even in a world where Weak Realism holds true, the EACP is not stronger than Causality, and is thus weaker than Locality.
 \textbf{Q.E.D.}

\medskip
\noindent
\textbf{Remark 2.} \emph{The first part of 
the second
proof of the EACP vs Locality Lemma tells us that the negation of the EACP in a world without augmentation permits signaling, which is enough to prove the EACP vs Locality Lemma in such a world.  Unfortunately, this first part does not exactly tell us that the EACP is trivially true as just being Causality since the EACP might still fail (only) in the limbo of entities that are well defined only because realism in some form holds true. Since such a world could be the actual world (although the final conclusion will be to contrary), we had to also consider a world accepting Weak Realism in the third proof. While a formal proof that the EACP is not Locality is achieved by restricting to a world with no realism at the microscopic level, this would not prove the EACP vs Locality Lemma as long as Weak Realism has not been proved wrong and we have to cover both cases in order to not fall in a circular argument. 
See also Remark 3 in subsection 3.3.}
  
\bigskip
\noindent
\textbf{3.3:}  \textbf{The No-Correlation Lemma.}

\smallskip
\noindent
{\bf No Correlation Lemma.} \emph{Assuming the EACP, if the oriented axes $a_{\mathcal E}$ and $a_{\mathcal E'}$ are orthogonal to each other, then the sequences $\mathcal E$  and $\mathcal E'$ are not correlated,} and more precisely we have:
\[
(\circ)\quad  \langle\mathcal E ,   \mathcal E'  \rangle=0 \quad\textrm{or equivalently}\quad Prob(\mathcal E_i = \mathcal E'_i)=\frac{1}{2}\,.
\] 
or more generally \[
\liminf_{N\to\infty}\frac{1}{N}[\mathcal E_{i+1}\cdot \mathcal E'_{i+1}
+\mathcal E_{i+N}\cdot \mathcal E'_{i+N}\leq 0\,,
\]
and 
\[
\limsup_{N\to\infty}\frac{1}{N}[\mathcal E_{i+1}\cdot \mathcal E'_{i+1}
+\mathcal E_{i+N}\cdot \mathcal E'_{i+N}\geq 0\,.
\]

\noindent
\emph{Proof of the No Correlation Lemma.}

We will first prove an auxiliary result (the \emph{Restricted No Correlation Lemma}) that corresponds to about the same statement, but when no measurement is made on the $P$ side. Then we will show that if measuring on the $P$ side would change the correlation, we could in principle have super-luminal message transmission for some entity that would have access to all quantities, measured or existing in virtue of Weak Realism. Next we will propose one protocol, \emph{the $\langle \mathcal E,\mathcal E'\rangle$ value Protocol} that lets one actually check that no change is made on the correlation  $\langle \mathcal E,\mathcal E'\rangle$ when a measurement is made on the $P$ side, leading to the lemma that we call \emph{Non-Locality permits signaling after all} (a result that clearly has an independent value and whose name comes from the fact that many have proven that affecting the value of  $\langle \mathcal E,\mathcal P\rangle$ (for instance: anyway the value of a correlation between values one on the $E$ side and one on the $P$ side) by an effect of Non-Locality does not permit  Super-Luminal message transmission.

\smallskip
\noindent
{\bf Restricted No Correlation Lemma.} \emph{Assuming the EACP and assuming further that no measurement is made on the $P$ side, if the oriented axes $a_{\mathcal E}$ and $a_{\mathcal E'}$ are orthogonal to each other, then the sequences $\mathcal E$  and $\mathcal E'$ are not correlated,} and more precisely we have:
\[
(R-\circ)\quad  \langle\mathcal E ,   \mathcal E'  \rangle=0 \quad\textrm{or equivalently}\quad Prob(\mathcal E_i = \mathcal E'_i)=\frac{1}{2}\,.
\] 
or more generally \[
\liminf_{N\to\infty}\frac{1}{N}[\mathcal E_{i+1}\cdot \mathcal E'_{i+1}
+\mathcal E_{i+N}\cdot \mathcal E'_{i+N}\leq 0\,,
\]
and 
\[
\limsup_{N\to\infty}\frac{1}{N}[\mathcal E_{i+1}\cdot \mathcal E'_{i+1}
+\mathcal E_{i+N}\cdot \mathcal E'_{i+N}\geq 0\,.
\]

\noindent
\emph{Proof of the Restricted No Correlation Lemma.}
Using the EACP and the conclusions that can readily be deduced from it, and more precisely the fact that ``the three sequences $\mathcal E$, $\mathcal E'$, and $\mathcal P$ involved in equation (\ref{Sica_3p_4})  are well defined",  we notice that, if furthermore no measurement is made on the $P$ side, then  only the orientation of the angle  $< a_{\mathcal E};  a_{\mathcal E'}>$ at $E$ could matter for an $E$-$P$ observer, so that $(\circ)$ follows from invariance under Parity without assuming Locality.  We next provide some details that some readers may prefer to avoid.

To see the role of Parity, we introduce the further oriented axis $ a_{\mathcal E''}$ that is parallel to $ a_{\mathcal E'}$ but with the opposite orientation.  This is the (only) oriented axis to which would correspond the sequence $\mathcal E''$ such that  $\mathcal E''_i\equiv -\mathcal E'_i$.  Since 
\begin{equation}\label{Sum2One}
Prob(\mathcal E_i = \mathcal E'_i)+Prob(\mathcal E_i = \mathcal E''_i)=1\,,
\end{equation}
it only remains to prove that these two probabilities are equal to each other.  We use here sequences whose values are possibly unknown (and indeed forever inaccessible to our knowledge), but that are  known to be well defined as we have recalled to begin this proof:

- One of these sequences, $\mathcal E$, is known by direct measurement, 

- The other sequence, $\mathcal E'$,  can be inferred to be well defined, even if unknown, by an $E$-$P$ observer on the basis of Quantum Mechanics augmented by  Weak Realism. 

\noindent
Since the angles $< a_{\mathcal E};  a_{\mathcal E'}>$ and $< a_{\mathcal E''};  a_{\mathcal E}>$ are equal, using the EACP, the only thing that could generate an inequality between $Prob(\mathcal E_i = \mathcal E'_i)$ and $Prob(\mathcal E_i = \mathcal E''_i)$ for an $E$-$P$ observer is the difference in the orientations of the angles $< a_{\mathcal E};  a_{\mathcal E'}>$ and $< a_{\mathcal E};  a_{\mathcal E''}>$.  Equality thus follows from Parity invariance if one assumes convergence of the means that define the correlations that are of intetrest to us.  Otherwise, in full generality, the argument that we have given proves that
\[
\liminf_{N\to\infty}\frac{1}{N}[\mathcal E_{i+1}\cdot \mathcal E'_{i+1}
+\mathcal E_{i+N}\cdot \mathcal E'_{i+N}\leq 0\,,
\]
and 
\[
\limsup_{N\to\infty}\frac{1}{N}[\mathcal E_{i+1}\cdot \mathcal E'_{i+1}
+\mathcal E_{i+N}\cdot \mathcal E'_{i+N}\geq 0\,.
\]

  \textbf{Q.E.D.}

Coming back to the full No Correlation Lemma, we see that the proof above cannot work if one does not assume Locality as the possible dependence of $E'$ upon the choice of $P$ breaks the symmetry about the bisector of the angle  $< a_{\mathcal E};  a_{\mathcal E'}>$. This is why we need to deal with the fact that some measurement will indeed made on the $P$ side. To do that, as we indicated before we will prove the \emph{Non-Locality permits signaling after all} Lemma, which in turn will need an ingredient of intrinsic value, the $\langle \mathcal E,\mathcal E'\rangle$ value Protocol, that we define and explain how to use next.

We notice that an entity able to access the values that exist in virtue of Weak Realism besides the values that can be measured would be able to send messages faster that light if the correlation would be changed when measurements are made on the $P$ side. The proof goes along the same way as what will be detailed next without assuming that one access the values assumed to exist as a result of Weak Realism.  We note however that, even without the results to come on superluminal signaling using observed quantities, we could terminate a (somewhat philosophically questionable prooof of the No-Correlation Lemma, from which a Bell Theorem follows as will be explained below. 

We shall formulate the following protocol to distinguish sources with correlations zero or significantly non-zero: any two correlation values can be treated the same way, the cost in time an precision having to be adjusted to mange correlation values closer than what we have to consider to reach our main conclusions.

\noindent
\emph{The $\langle \mathcal E,\mathcal E'\rangle$ value Protocol.} Consider two sources $S_0$ and $S_1$ that at discrete times $t_1, t_2,\dots , t_n,\dots$ produce either a $S_i(t_n)=-1$ or a $S_i(t_n)=1$ so that the processes of each source is Markovian (\textit {i.e.} memory-less) with both outputs equally probable on each source considered on it own. Assume that the correlation 
\[
\lim_{N\to\infty}\frac{1}{N}[(S_0(t_k)\cdot S_1(t_{k+1})
+\dots +(S_0(t_{k+N})\cdot S_1(t_{k+N})=0
\]
with some setting $\textbf{S}_a$ and that the quantity  
\[
\frac{1}{N}[(S_0(t_k)\cdot S_1(t_{k+1})
+\dots +(S_0(t_{k+N})\cdot S_1(t_{k+N})
\]
fails to converge to zero some other setting $\textbf{S}_b$. Then this difference of behavior can be detected
by using only one of $S_0$ and $S_1$ at each time step, using the following method:

- First one chooses one big number $Q$, say $Q=1001$ and another number $P$ not too small when compared to $Q$ but nevertheless significantly smaller than $Q$, say $P=100$.  One then choose one of the two source, $S_i$, and compute 
\[
\sigma_{(1,1)}=S_i(t_k)+S_i(t_{k+1})+\dots+ S_i(t_{k+Q}).
\]
If $|\sigma<P$, then one computes
\[
\sigma_{(1,2)}=S_i(t_k+Q+1)+S_i(t_{k+Q+1}+\dots+ S_i(t_{k+2\cdot Q}).
\]
and so on until one arrives at some first $\sigma_{(1,j)}$ such that  $|\sigma_{(1,j)}|\geq P$.
Then one computes  
\[
\sigma'_{(1,j)}=S_{1-i}(t_{j\cdot k+1})+S_{1-i}(t_{j\cdot k+2})+\dots+ S_{1-i}(t_{(j+1)\cdot k}).
\]
and one sets:
\begin{itemize}
\item $ v_{(1,j)}=-1$ if $\sigma'_{(1,j)}\leq-P\,,$
\item  $v_{(1,j)}=0$ if $|\sigma'_{(1,j)}|<P\,,$
\item $ v_{(1,j)}=+1$ if $\sigma'_{(1,j)}\geq P\,.$
\end{itemize}
One proceed the same way some very large number $L$ of times. Typically, we used $L=4 500 00$ when using $Q=1000$ and $P=100$ to get very decisive differences between correlations -1, 0, and 1 between the isochrone signals of $S_0$ and $S_1$.   Once the number $L$ is chosen (but it can easily be enlarged if one wants to differentiate between two correlations values that are closer to each other than anything one would have studied before), for $l=1,2,\dots, L$, one computes the sums of the $ v_{(l,j)}$'s conditioned respectively by $\sigma_{(l,j)}\leq -P$ and $\sigma_{(l,j)}\geq P$, altogether, after dividing the conditioned sums by $L$, four numbers, say  $V_{(l,j)}(+,+), V_{(l,j)}(+,-), V_{(l,j)}(-,+), V_{(l,j)}(-,-)$
that are all between 0 and 1, and that form together a signature of the correlation of the two sources. The finite size aspects that must be handled to turn what has been described here into a practical algorithm are standard and the details, that can take many possible forms, are left to the taste and care of the reader.

The fact that the quadruplet $(V_{(l,j)}(+,+), V_{(l,j)}(+,-), V_{(l,j)}(-,+), V_{(l,j)}(-,-))$ is sensitive to correlation permits signaling out of data that come at each time from one of the two sources, that can at each step be chosen by the experimentalist. Since what we want is detect a difference with zero correlation, the fact that the sequence on the $P$ side depends on the choice made on the other side is irrelevant. From there come several consequences:

- Non-locality, if generating changes able to avoid the contradiction generated by Bell's Theorem, would permit super-luminal signaling, hence would be impossible if Special Relativity is to be preserved.

- The non-exiatence of super-luminal signaling generated that way, \textit{i.e.,} the independence of the correlation $\langle \mathcal E,\mathcal E'\rangle$ upon a measurement being made on the other side but assuming Weak Realism, can be checked by direct experiment. That experiment, without the Weak Realism Assumption, would show that the restriction of the sources $S_0$ and $S_1$ signal (\textit{i.e.,} the measurements of the projections of the spins along $E$ or $E'$ behave like restrictions of uncorrelated signals.

- A Bell type theorem can be proven while assuming the EACP instead of Locality (although as we have seen, the EACP implies that one of Locality or the possibility of super-luminal signaling has to hold true).

This finishes the proof of the No-Correlation Lemma and concludes as well the proof of the two other important results that were brought to bare:

\noindent
-  the capability of the proposed protocol to compare different values of the correlation  $\langle \mathcal E,\mathcal E'\rangle$, 

\noindent
- and the application of that protocol to prove that Non-Local behavior of the form ``measuring on the $P$ side change the value $\langle \mathcal E,\mathcal E'\rangle$" would permit super-luminal message transmission, with a mean to verify experimentally that no change in the value of $\langle \mathcal E,\mathcal E'\rangle$" happens in a way that would indeed permit super-luminal message transmission.   \textbf{Q.E.D.}

\bigskip
\noindent
\textbf{3.4:}  \textbf{A Bell's Theorem with no Locality assumption.}
In this subsection, we formulate and prove the following other main result of this paper beside the protocol and its applications mentioned above.  More precisely, the tools and fact that we have assembled let us formulate and prove the following result:

\smallskip
\noindent
\textbf{New Bell's Theorem:} \emph{Assuming  Weak Realism and the EACP, we can use the triplet
of angles $(\theta_\mathcal P,\theta_\mathcal E,\theta_{\mathcal E'})= (0,\frac{3\pi}{4}, \frac{-3\pi}{4})$
that corresponds to the triplet of correlations
$( \langle  \mathcal P , \mathcal E\rangle,\langle\mathcal E' ,\mathcal P\rangle,  \langle\mathcal E , \mathcal E'\rangle)=(\frac{\sqrt{2}}{2}, \frac{\sqrt{2}}{2}, 0)\,$  
to generate a contradiction using the $V3$ version of Bell's Inequalities.}

\smallskip
\emph{Proof of the New Bell's Theorem.}  Again we assume the EACP and we use
equation (\ref{Sica_3p_4}) as the inequality to be falsified.
 
 \smallskip
- 1) After measurements are made using $P$, a $P$-$E$ observer obtains  that:

\noindent
- e1) $\langle \mathcal P,\mathcal E\rangle=\frac{\sqrt{2}}{2}$, \textit{i.e.,}  $\langle \mathcal P,\mathcal E\rangle\approx 0.7$ by Quantum Mechanics, or by direct observation after measurements are also made using $E$,

\noindent
 - e2) $\langle \mathcal P,\mathcal E'\rangle= \frac{\sqrt{2}}{2}$, \textit{i.e.,} $\langle \mathcal P,\mathcal E'\rangle\approx 0.7$ by Quantum Mechanics augmented by  Weak Realism.
 
 The deductions made in e1) and e2) using Quantum Mechanics augmented by Weak Realism go as follows:  By Wave Packet Reduction (for instance), the spin state of second particle (the particle on the $E$ side) becomes 
\begin{equation}\label{Collapse}
\Psi(x_2)=|\mathcal P_i \rangle _1\otimes| -\mathcal P_i \rangle_2 
\end{equation}
along the oriented axis along which the sequence $\mathcal P$ is measured, as soon as the measurement of $\mathcal P_i$ is made on the $P$ side.  Hence the second particle gets into a spin state prepared to be $ |-\mathcal P_i \rangle $ along that oriented axis
(as revealed by using the information obtained on the $P$ side) so that both of the two correlations $\langle\mathcal P,\mathcal E\rangle$ and $\langle \mathcal P,\mathcal E'\rangle$ are equal to $\frac{\sqrt{2}}{2}$ (about $0.7$) by a simple application of the twisted Malus law as we have recalled it, under Convention 2 and the EACP, as we saw in subsection 2.2.

 \smallskip
- 2) An $E$-$P$ observer infers that: 

\noindent
 - e3) $Prob (\mathcal E_i=\mathcal E'_i)= 0.5$ (\textit{i.e.,} $\langle \mathcal E, \mathcal E'\rangle =0$) on the $E$ side by the No Correlation Lemma. 

\smallskip
Assembling the conclusions  e1), e2), and e3) from the two (strongly) asynchronous frames (\textit{e.g.,} in the Lorentz frame of the experiment since the outcomes cannot change according to the Lorentz frame by relativistic invariance of observable events) one obtains  the expected triplet evaluation for the three correlations: $(\frac{\sqrt{2}}{2}, \frac{\sqrt{2}}{2}, 0)$.  Together with equation (\ref{Sica_3p_4})  (the $V3$ version of Bell's Inequalities), this evaluation provides us with the impossible inequality $1.4\leq 1$, or more precisely $\sqrt{2}\leq 1$ as when we examined equation (\ref{Sica_3p_4}) while assuming Locality in subsection 2.2.  This concludes the proof of the New Bell's Theorem.  \textbf{Q.E.D.}

\medskip
Our New Bell's Theorem admits the following immediate corollary that we will use as our main conclusion:

\medskip
\noindent
\textbf{ Conclusive Corollary:} \emph{ Weak Realism is the \textbf{only} possible cause of contradiction common to \textbf{all} the versions of Bell's Theorem and some of the Bell's type contradictions cannot be solved by assuming Non-Locality. Thus Non-Locality is not needed (in some circles, one would say that Non-Locality can be disposed of using Occam's razor).} 

\medskip
\noindent
\textbf{Remark 3.} \emph{As we saw, without (Weak) Realism any violation of the EACP is a violation of Causality since then the EACP is one of the expressions of Causality.  In order for the violation of the EACP to not be a violation of Causality, one would have to accept that, with probability one, the negation of the EACP has effect \emph{only} on values of observable that are linked to (Weak) Realism.  We do find that unacceptable and we thus consider that the new Bell Theorem condemns Weak Realism. This is not a proof and possibly no actual proof can be given to help us decide between keeping  Weak Realism and keeping the EACP but this situation is more frequent than one might think.  Mathematics and not physics is the realm of ``proofs": there has always been some opinions lurking behind the way we apprehend the laws of Physics. 
Some would say that as soon as one uses  Weak Realism, one steps into Metaphysics anyway.  However we point out that  Weak Realism violates the spirit if not the letter of the Uncertainty Principle \cite{Heisenberg} or at least its time-reversed version \cite{ETP, Tresser2}, and is in particular rejected by the Copenhagen Interpretation of Quantum Mechanics (which consequently would have to statute that Bell'Theorem has nothing to say about Quantum Mechanics).  On the other hand, invoking  Weak Realism, even if bad, is probably not as bad as accepting that the EACP is false, yet the present
work shows that accepting the mildly unacceptable Weak Realism implies accepting the quite unacceptable violation of the EACP.}

\medskip
\noindent
\textbf{Remark 4.} \emph{The well known Hidden Variable theory of de Broglie and Bohm \cite {deBroglie19271956}, \cite{Bohm1952} is both Non-Local and Realist, yet it avoids the contradictions that constitutes BellÕs theorem.  In fact it avoids these contradictions  precisely because Non-Locality prevents the Bell's inequality from making sense. This statement on the de Broglie - Bohm Theory (dBBT) is not a contradiction to our Conclusive Corollary about Weak Realism and Locality nor more generally to the theses defended here.  Indeed, the dBBT is massively not Lorentz invariant, way beyond the special setting for Bell's theory, and apparently irreducibly so: in any Bohmian quantum theory the quantum equilibrium distribution $|\Psi|^2$ cannot simultaneously be realized in all Lorentz frames of reference.  To the contrary, Quantum Mechanics can be viewed as a non-relativistic approximation to relativistic Quantum Field Theory in the limit when Classical Mechanics is a good approximation to Special Relativity. The dBBT, or Bohmian mechanics, the version re-discovered and extended by Bohm, is thus false or at least considered as false by most physicists, even if it can serve pedagogically as advocated by Bell (this opinion of Bell, who defended Bohmian Mechanics, may not be shared by those who consider Non-Realism as an essential ingredient of microphysics).  Indeed, some  Bohmian physicists still hope for a version of Bohmina Mechanics that could be acceptable by the profession, but it should be noted that (many) Bohmian physicists take as a strong argument in favor of Bohmian mechanics \emph{the false fact that Bell's Theorem and Alain Aspect's experiments prove Quantum Mechanics to be a non-local theory}. Indeed, in some sense, Bell's Theorem can be considered as the proof that \emph{the Non-Local character of the dBBT was irreducible among Hidden Variable Theories and more generally among Realist Theories}.  To the contrary since it assumes (Weak or stronger) Realism, hence forces us out of Quantum Mechanics (into Bohmian Mechanics or some weak form of it), \emph{Bell's Theory and related experiments have nothing to say about the Locality or non-Locality of Quantum Mechanics itself}, only statements about some augmentations of Quantum Mechanics. This being said, Bell and others consider that the complete nature of Quantum Mechanics can be trusted so that the EPR paper proves Quantum Mechanics itself to be non local, but here is not the place for a debate on the History of Sciences.}

%
\medskip
Like in the case of the EPRB entanglement (see \textit{e.g.,} \cite{AspectEtAl1982}), experiments have been done on the GHZ entanglement (see \textit{e.g.,}  \cite{BPWZ1999}). Some statistical analysis done on the GHZ experiments, which use the less than perfect performance of the captors,  such as \cite{SzaboFine2002}  and \cite{HessPhilipp2004} and the critical papers responding to these attacks will not be considered here, and neither will other entanglements such as in \cite{Hardy1993}.

\bigskip

\begin{acknowledgments}
\textbf{Acknowledgments:} 
Many people have helped me in this enterprise, by vicious attacks, constructive questions, encouragements,  patient listening, and often friendship: Y. Avron, M. le Bellac, O. Cohen, P. Coullet, D. Greenberger, R. Griffiths, G. t'Hooft, A. Mann, D. Mermin, D. Ostrowsky,  I. Pitowsky, Y. Pomeau, O. Regev, M. Revzen, T. Sleator, J. Tredicce,  L. Vaidman and many more.  Some people could have as well been on the front page but declined.  I cannot find the words to thank Arthur Fine, Richard Friedberg, Pierre Hohenberg, Larry Horwitz, Marco Martens, and Edward Spiegel for their patience, strong but important and legitimate critics, advices, and encouragements.  Meeting with Dan Greenberger, and his kind interest in my work, made me realize the GHZ part of this paper. At last I thank a referee, in particular for pointing out to me the work of Louis Sica \cite{Sica1, Sica2}.  This referee also mentioned a very interesting paper of Hardy \cite{Hardy1992} that we hope to discuss at length and from many aspects in work to come.
\end{acknowledgments}

%
%
%
%

\end{document}